\documentclass[pra,aps,twocolumn,amsmath,amssymb]{revtex4}
\usepackage{graphicx}
\pdfoutput=1
\usepackage{amsmath}
\usepackage{color}
\usepackage{float}
\setcitestyle{super}
\usepackage[english]{babel}
\usepackage{color,soul}

\begin{document}
\title{Low-energy excitations and non-BCS superconductivity in Nb$_x$-Bi$_2$Se$_3$}

\author{Anshu Sirohi$^1$, Shekhar Das$^1$, Prakriti Neha$^2$, Karn S. Jat$^2$, Satyabrata Patnaik$^2$}

\author{Goutam Sheet$^1$}
\email{goutam@iisermohali.ac.in}

\affiliation{$^1$Department of Physical Sciences, Indian Institute of Science Education and Research(IISER) Mohali, Sector 81, S. A. S. Nagar, Manauli, PO: 140306, India.}

\affiliation{$^2$School of Physical Sciences, Jawaharlal Nehru University, New Delhi, PO: 110067, India}
\date{\today}

\begin{abstract}

\textbf{When certain elemental metals like Cu, Sr and Nb are intercalated between the layers of Bi$_2$Se$_3$, a topological insulator, the intercalated systems superconduct with critical temperatures around 3 K. Naturally, in all these cases, the possibility of topological superconductivity was suggested and explored. However, in cases of Cu and Sr intercalated systems, the low-temperature scanning tunneling microscopy (STM) experiments revealed fully formed gaps where no signature of low-energy states, a requisite for topological superconductivity, was found. Here, through STM spectroscopy down to 400 mK we show that in Nb$_x$-Bi$_2$Se$_3$ ($x$ = 0.25), the spectra deviate from a BCS-like behavior and the spectral weight at low-bias is large. Our observations are consistent with the idea that the order parameter of Nb$_x$-Bi$_2$Se$_3$ is nodal. Therefore, our results conclude that compared to other members of the family, Nb$_x$-Bi$_2$Se$_3$ has a stronger possibility of being a topological superconductor.}

\end{abstract}

\maketitle

Majorana fermions,\cite{Kane, YAndo, MSato, SDS1, SDS2, SDS3, SDS4, Leo1} the hitherto elusive particles that were predicted to exist in the form of elementary particles described by the Dirac equation,\cite{PAMDirac} might appear as collective excitations in condensed matter systems.\cite{SDS1, Steven} In a condensed matter system, in order for a collective excitation to be considered as a Majorana fermion, it must satisfy two conditions: (i) it must obey Dirac equation.\cite{PAMDirac, Steven} This can be verified by the observation of a linear dispersion near a band crossing point. (ii) The excitation should be its own antiparticle.\cite{Steven} Both of these conditions  are  expected to be satisfied in topological superconductors.\cite{Ludwig, Zhang1, Zhang2} Due to the topological nature and the bulk-boundary correspondence,  gapless excitations described by Dirac equation are supported on the boundaries of topological superconductors.\cite{Ludwig, Zhang1, Zhang2} Hence, the first condition is satisfied. The second condition is satisfied because in a superconductor, the electron and hole wave functions are mixed. This  helps a  superconductor achieve particle-hole symmetry leading to the gapless excitations on the surface being their own anti-particles.\cite{Tinkham} Motivated by this idea, efforts have been made to realize condensed matter systems where topologically non-trivial bands might coexist with superconductivity.\cite{Liu} The simplest approach towards that, in principle, would be to induce superconductivity in topological insulators. In reality, when certain elemental metals like Cu, Sr and Nb were intercalated between the interleaved planes of the most celebrated topological insulator Bi$_2$Se$_3$,\cite{Bi2Se3_Zhang, Bi2Se3_Hasan} the resultant systems showed superconductivity with critical temperatures around 3 K in all cases.\cite{CuBi2Se3_Cava, CuBi2Se3_Ando, SrBi2Se3_Pneha, SrBi2Se3_Han, MPSmiley2} In cases of Cu$_x$-Bi$_2$Se$_3$ and Sr$_x$-Bi$_2$Se$_3$, STM experiments revealed a fully formed superconducting gap without any signature of low-energy excitations.\cite{CuBi2Se3_Ando, CuBi2Se3_Levy_STM, SrBi2Se3_STM} Observations of fully gaped superconducting gap cast doubt on whether the superconductivity is conventional (BCS-like) or unconventional.\cite{BCS} Theories and experimental observations on these compounds have also been conflicting, obscuring the realization of topological superconductivity to date.\cite{CuBi2Se3_TSC1, CuBi2Se3_TSC2, CuBi2Se3_TSC3, CuBi2Se3_TSC4} Here, we show that the third candidate in this family, Nb$_x$-Bi$_2$Se$_3$, shows low-energy excitations where the superconducting gap is seen to not fully form down to 385 mK, almost one tenth of $T_c$. Furthermore, the tunneling spectra deviate from the BCS predictions indicating the presence of a complex or mixed angular momentum symmetry in the superconducting order parameter.

In the past, it was shown that in Nb$_x$-Bi$_2$Se$_3$, when the superconducting phase is realized, the topologically non-trivial properties are also preserved.\cite{NBS_Qiu}. An exotic nematic order in the superconducting phase of Nb$_x$-Bi$_2$Se$_3$ was also reported.\cite{NBS_Asaba, NBS_Shen} From heat capacity measurements it was suggested that this system has a nodeless superconducting gap consistent with odd-parity p-wave pairing.\cite{NBS_Lawson}  More recently, based on proton-irradiated samples of Nb$_x$-Bi$_2$Se$_3$ it was claimed that the superconductivity in this system arises from odd frequency pairing where the order parameter also has symmetry protected nodes in the momentum space.\cite{MPSmiley1} However, the nature of the superconducting gap has not been investigated by high energy-resolution scanning tunneling spectroscopy in the superconducting state.

High quality single crystals of Nb$_x$-Bi$_2$Se$_3$ were used for the measurements presented in this paper. The crystals were grown by a modified bridgeman technique. The crystals were extensively characterized by structural and magnetic measurements and as shown in the magnetization vs. temperature data in Figure 1(a), a superconducting transition at 3.5 K was observed. 
\begin{figure}[h!]
		\includegraphics[width=0.5\textwidth]{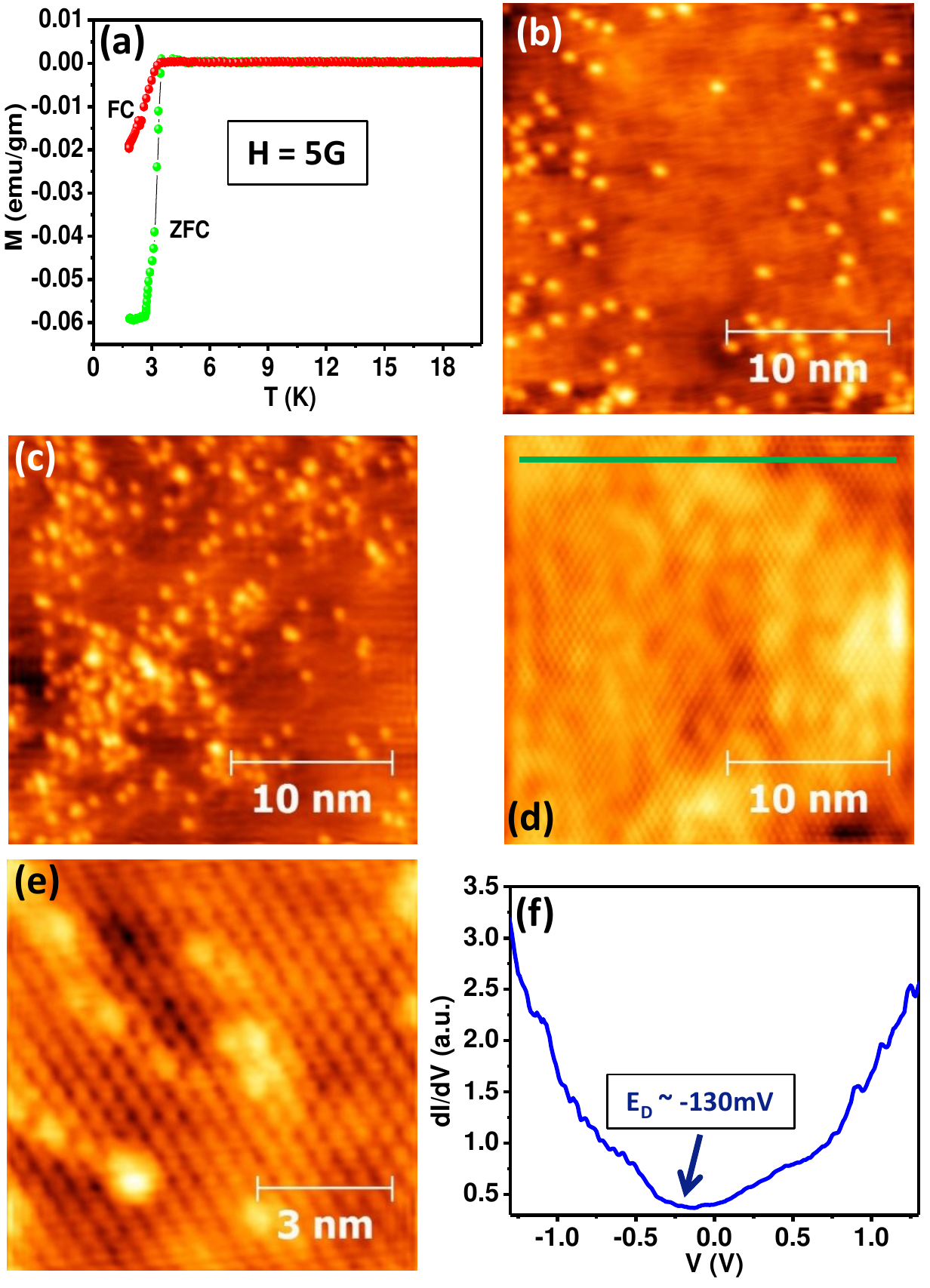}
		\caption{(a) M vs. T ZFC (Green curve) and FC (Red curve) showing superconducting transition at 3.5 K.  STM topographs of 25 nm x 25 nm area showing Nb clusters with (b) lower density and (c) higher density recorded at 18 K (V$_b$ = 1 V, I$_t$ = 110 pA). (d) Topograph of a cluster-free area showing atoms with an underlying bright/dark contrast recorded at 400 mK (V$_b$ = 10 mV, I$_t$ = 5 pA). (e) Topograph of an area with atomically resolved background and Nb clusters recorded at 2K (V$_b$ = 10 mV, I$_t$ = 1.2 pA). (f) Tunneling spectroscopy at 4K showing a Dirac-like dispersion.}	
	\label{Figure 4}
\end{figure}

The STM and STS experiments were carried out in an ultra-high vacuum (UHV) cryostat working down to 380 mK (Unisoku system with RHK R9 controller). First a single crystal of Nb$_x$-Bi$_2$Se$_3$ was mounted in a low-temperature cleaving stage housed in the exchange chamber of the system where the crystal was cleaved by an $in-situ$ cleaver at 77 K in UHV ($10^{-11}$mbar). After cleaving, the crystal was immediately transferred by an UHV manipulator to the scanning stage hanging with three metal springs at low-temperature. This process minimized the possibility of contamination and/or modification of the pristine surface exposed by UHV cleaving. In Figure 1(b) we show an STM topograph of an area of 20 nm x 20nm.  In the image we observe disc shaped bright objects with a flat background. These bright objects are the defect states due to Nb atoms/clusters that the system has been intercalated with.\cite{NBS_Qiu} These objects are randomly distributed over the surface and their concentration is different in different regions of the crystal. In Figure 1(c) we show a different area where the density of the Nb clusters is considerably larger. However, we did not find any difference in superconducting properties between regions with different density of Nb clusters. This observation is different from the STM experiments on Sr$_x$-Bi$_2$Se$_3$,\cite{SrBi2Se3_STM} where the superconducting properties varied dramatically from region to region on the surface and for certain regions with lower concentration of Sr clusters, superconductivity was absent.\cite{SrBi2Se3_STM}
\begin{figure}[h!]
		\includegraphics[width=0.5\textwidth]{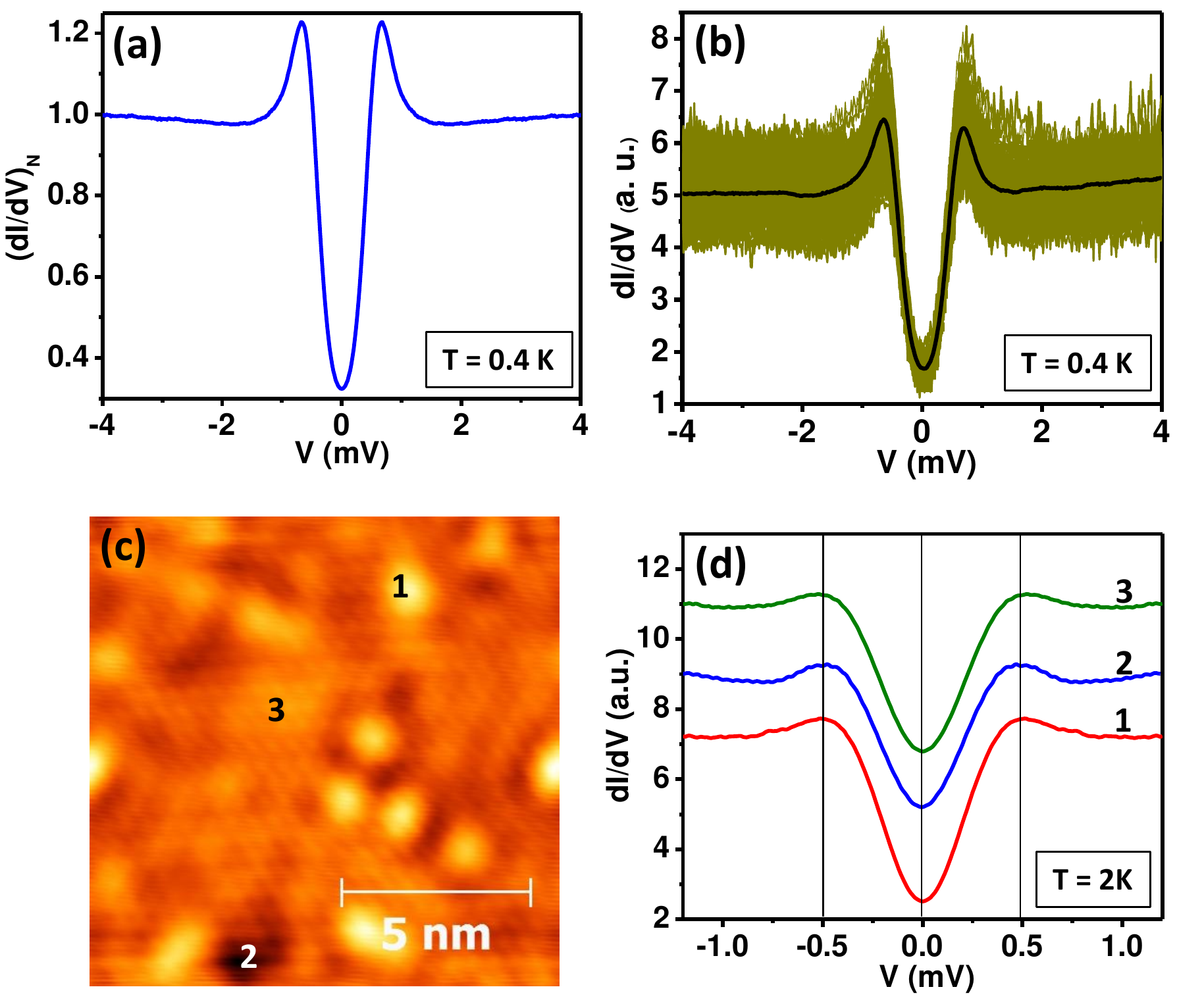}
		\caption{(a) A representative tunneling spectrum at 400 mK showing superconducting gap. (b) The spectra at 400 mK at multiple points along the green line shown in figure 1(d). The black line is the average of all the spectra. (c) Topograph of a 12 nm x 12 nm area showing atoms, background dark/bright contrast and Nb clusters. (d) Tunneling spectra recorded at 2K at three points $``1"$, $``2"$, $``3"$ as shown in (c).}	
	\label{Figure 4}
\end{figure}

We also found large areas where the Nb clusters are not resolved. In Figure 1(d) we show the image of a 25 nm x 25 nm area captured at an STM bias of 10 mV where the disc-shaped clusters are not seen. Under this condition, the atoms on the surface of the crystal become clearly visible. A visual inspection of the atomically resolved image reveals alternate bright and dark regions in the background. In order to confirm whether this contrast might arise due to the Nb clusters in a layer below the top surface, we found an area where the clusters as well as the well resolved atoms are seen in the same topograph (Figure 1(e)). It is seen that the bright and dark contrast displays no correlation with the position of the clusters. Therefore, the contrast might originate from an intrinsic property of the system. A qualitative comparison of the STM image with that obtained on the compensated topological insulator BiSbTeSe$_2$,\cite{BSTS} it can be argued that the contrast might be due to puddling of electron rich and hole rich areas on the surface. Scanning tunneling spectroscopy (STS) at 4 K at different points on the surface reveals a ``Dirac cone" shape with a possible Dirac point only $\sim$ 130 meV below the Fermi energy (Figure 1(f)). The existence of a Dirac point so close to the Fermi energy indicates that the Dirac character of the dispersion may extend beyond the Fermi energy thereby supporting the idea of charge puddling in this case. This also indicates that the helical states of the surface may participate in superconducting pairing, fulfilling the requirement for topological superconductivity here.

\begin{figure}[h!]
	\centering
		\includegraphics[width=0.5\textwidth]{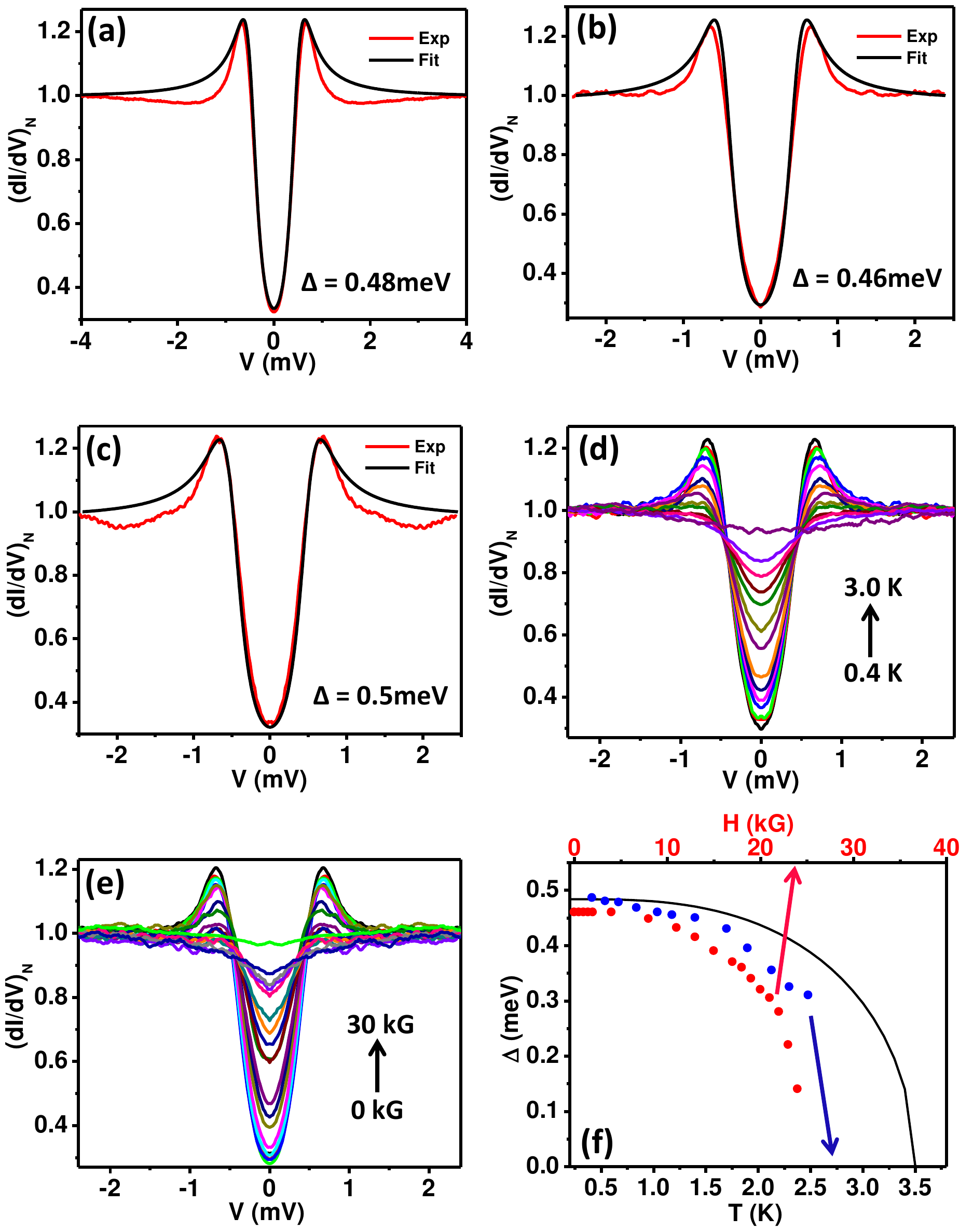}
		\caption{(a, b, c) Three representative spectra recorded at 400 mK with fits using an extended BCS theory. (d) Temperature dependence of a typical spectrum. (e) Magnetic field dependence of a typical spectrum. (f) Temperature (blue) and magnetic filed (red) dependence of $\Delta$ as extracted from fitting of the spectra in (d) and (e). The black line shows the BCS prediction of the temperature dependence of $\Delta$.}	
	\label{Figure 4}
\end{figure}
Now we focus on the tunneling spectroscopy on the surface of Nb$_x$-Bi$_2$Se$_3$ below the superconducting critical temperature. In Figure 2(a) we show a typical STS spectrum obtained at 400 mK. Two coherence peaks followed by a low-bias dip indicating the formation of a superconducting gap is clearly visible in the spectrum. However, unlike a typical spectrum with a fully opened BCS gap, in this case, the quasiparticle density of states at zero energy is large. To note, in cases of Cu$_x$Bi$_2$Se$_3$\cite{CuBi2Se3_Levy_STM} and Sr$_x$Bi$_2$Se$_3$,\cite{SrBi2Se3_STM} a fully formed superconducting gap with $dI/dV$ = 0 at zero bias was seen. Such spectral features without a fully formed gap are reproducibly observed at large number of points on the surface. To confirm this, in Figure 2(b) we plotted a large number of spectra obtained at different points on the green line shown in Figure 1(d). Along the line that cuts through dark, bright and very bright regions, as it is seen, the coherence peaks appeared at the same energy indicating the insensitivity of the superconducting phase to the underlying modulation of the local density of states. To confirm this unusual robustness of the superconducting phase, we selected an area (Figure 2(c)) where the Nb clusters as well as the dark-bright contrast with an atomically resolved background can be seen and performed local tunneling spectroscopy at certain special points with different brightness (points 1, 2 and 3, for example) at 2 K, very close to the $T_c$. As shown in Figure 2(d) the spectra obtained at points 1, 2 and 3 did not show any qualitative difference. It is also interesting that in all these points, superconducting spectral features not only exist but also show striking similarities with each other at a temperature close to $T_c$. This indicates that the $T_c$ is also uniform over the crystal surface, despite the presence of the underlying contrast.                  

\begin{figure}[h!]
	\centering
		\includegraphics[width=0.5\textwidth]{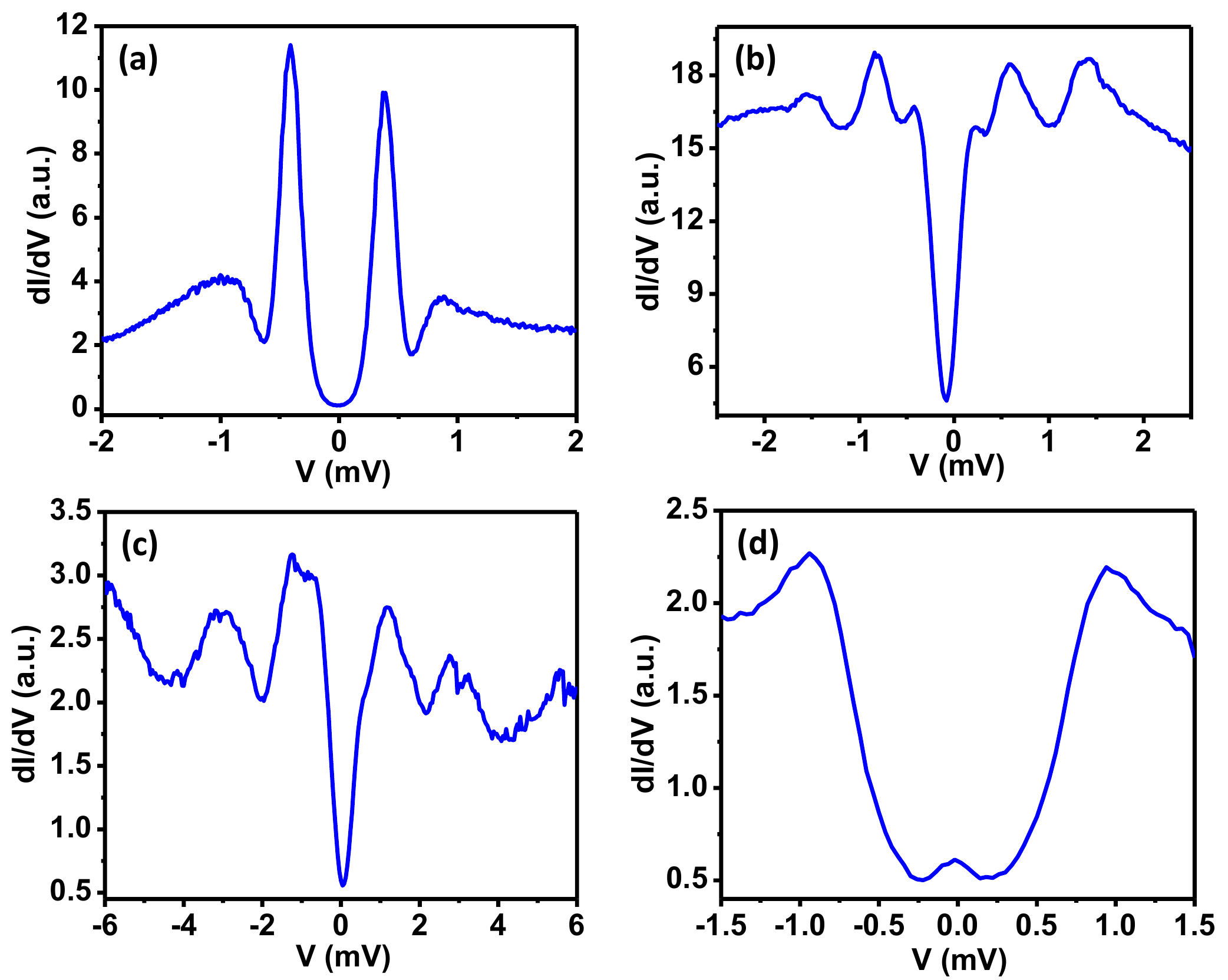}
		\caption{(a) A spectrum showing prominent dips above coherence peaks. (b,c) Two tunneling spectra showing multiple coherence peaks. (d) One spectrum showing zero-bias conductance peaks.}	
	\label{Figure 4}
\end{figure}

In Figure 3(a,b,c) we show three representative tunneling spectra (red lines) with best possible theoretical fits (black lines) using a model which relies on an extension of BCS theory.\cite{BCS}  Within this formalism we used  Dyne's formula $N_s(E) = Re\left(\frac{(E-i\Gamma)}{\sqrt{(E-i\Gamma)^2-\Delta^2}}\right)$, where $\Gamma$ is an effective broadening parameter incorporated to take care of slight broadening of the BCS density of states possibly due to finite quasi-particle life time.\cite{Dynes}The fitting parameters are the superconducting energy gap $\Delta$ and the broadening parameter $\Gamma$ which is included in the theory as a complex correction to the energy $E$. The non-zero conductance at $V$ = 0 is obtained by putting a relatively higher value of $\Gamma$ within this theory. Apart from the large density of states at low-energy, the experimental data deviate from the theoretical lines more dramatically just above the coherence peaks where the experimental spectra show a shallow dip. This feature is too common and reproducible in this system to be ignored as a moderate deviation from the theoretical fit. Therefore, considering the observed large low-bias density of states and the deviation of the experimental spectra from fits within the BCS theory,\cite{BCS} it can be concluded that the superconducting phase in Nb$_x$-Bi$_2$Se$_3$ has unconventional character. However, the possibility of a mixed angular momentum symmetry where an unconventional component coexists with an $s$-wave gap may also be possible. This is consistent with other experiments where indications of a superconducting gap with point nodes in Nb$_x$-Bi$_2$Se$_3$ was reported.\cite{MPSmiley2}  

The spectrum at 400 mK smoothly evolves with increasing magnetic field as for a type II superconductor and all the spectral features disappear at a critical field of 3 Tesla (Figure 3(e,f)). In order to further study the nature of superconductivity, we have performed the temperature dependence of the tunneling spectra (Figure 3(d)). The spectral features associated with superconductivity gradually decrease and they completely disappear at 2.8 K, slightly below the global $T_c$. Since the experimental data do not match with the conventional theory that we have used to fit them, evaluation of the exact temperature dependence of the gap ($\Delta$) is non-trivial. We have plotted $\Delta$ obtained from the best possible fitting for the spectra at different temperatures (Figure 3(f)). In the same panel we also show the expected behavior within BCS theory.\cite{BCS} The experimental temperature dependence of $\Delta$ follows the BCS-like dependence at very low temperatures but drifts away from that at higher temperatures.

While majority of the tunneling spectra obtained in the superconducting state of Nb$_x$-Bi$_2$Se$_3$ showed  deviation from the theoretical fits as shown in Figure 3(a,b,c), other spectral features were also often obtained at certain points on the surface for which the deviation from the theoretical spectra were more significant. Four such spectra are shown in Figure 4(a,b,c,d). The spectrum in Figure 4 (a) is qualitatively similar to the spectra presented in Figure 3 except for the appearance of prominent dips just above the coherence peaks. To note again, all the spectra in Figure 3 deviate from the fits due to the formation of shallow dips just above the coherence peaks. Such a dip structure is predicted to exist in the tunneling spectrum between a normal metal and a superconductor with chiral pairing of the superconducting order parameter.\cite{Chiralpwave_Tanaka1, Chiralpwave_Tanaka2} Hence, an $s + ip$ symmetry of the superconducting gap might be possible in this case. This is again consistent with earlier reports of the observation of nodal order parameter symmetry in Nb$_x$-Bi$_2$Se$_3$.\cite{MPSmiley2}  

For certain spectra (Figure 4(b,c)), multiple (coherence) peaks in $dI/dV$ spectra were observed. This might be due to the tip picking up a niobium atom from the surface which might lead to the formation of an S-I-S type of junction, where Josephson type tunneling may dominate the transport.\cite{CuBi2Se3_Levy_STM} The other type of spectra obtained on Nb$_x$-Bi$_2$Se$_3$ showed a prominent zero-bias conductance peak. Such a zero-bias conductance peak (ZBCP) might originate from a number of sources including magnetic impurities and unconventional or topological character of the superconducting phase. Though we have ruled out the existence of magnetic impurities in our crystals, based on our STM spectroscopic results alone it is not possible to conclude whether the ZBCP indeed originates from a topological character of the superconducting phase. However, the occasionally appearing spectra presented in Figure 4 provide further support for the unconventional (non-BCS) character of superconductivity in Nb$_x$-Bi$_2$Se$_3$.

In conclusion, from scanning tunneling microscopy and spectroscopy we have shown that the superconducting phase realized by intercalation of Nb in the topological insulator Bi$_2$Se$_3$ is not described by BCS theory. The surface of Nb$_x$-Bi$_2$Se$_3$ shows charge puddling indicating the topological dispersion exists in the energy scale of superconductivity. Unlike in case of the sister materials Cu$_x$-Bi$_2$Se$_3$ and Sr$_x$-Bi$_2$Se$_3$, in Nb$_x$-Bi$_2$Se$_3$ the superconducting energy gap does not form fully and the STM spectra showed evidence of low-energy excitations present down to a temperature $\sim$ 0.1 $T_c$.  

We thank Tanmoy Das for reading the manuscript critically and for his valuable comments. GS would like to acknowledge financial support from the research grant of Swarnajayanti fellowship awarded by the Department of Science and Technology (DST), Govt. of India under the grant number DST/SJF/PSA-01/2015-16.


\begin{thebibliography} {99}


\bibitem{Kane} C. L. Kane, and E. J. Mele, \textit{Phys. Rev. Lett.} \textbf{95}, 146802 (2005).

\bibitem{YAndo} M. Sato, and Y. Ando, \textit{Rep. Prog. Phys.} \textbf{80}, 076501 (2017).

\bibitem{MSato} M. Sato and S. Fujimoto, \textit{J. Phys. Soc. Jpn.} \textbf{85}, 072001 (2016).

\bibitem{SDS1} J. D. Sau, S. Tewari,R. M. Lutchyn, T. D. Stanescu, and S. D. Sarma, \textit{Phys. Rev. B.} \textbf{82}, 214509 (2010).

\bibitem{SDS2} R. M. Lutchyn, J. D. Sau, and S. D. Sarma, \textit{Phys. Rev. Lett.} \textbf{105}, 077001 (2010).

\bibitem{SDS3} T. D. Stanescu, R. M. Lutchyn, and S. D. Sarma, \textit{Phys. Rev. B.} \textbf{84}, 144522 (2011).

\bibitem{SDS4} S. D. Sarma, J. D. Sau, and T. D. Stanescu, \textit{Phys. Rev. B.} \textbf{86}, 220506(R) (2012).

\bibitem{Leo1} C. Beenakker, and L. Kouwenhoven, \textit{Nature Physics} \textbf{12}, 618 (2016).

\bibitem{PAMDirac}  P. A. M. Dirac, 1928, \textit{Proc. R. Soc. A}, \textbf{117}, 610, (1928).

\bibitem{Steven} S. R. Elliott and M. Franz, \textit{Rev. Mod. Phys.}, \textbf{87}, 137, (2015).

\bibitem{Ludwig} A. P. Schnyder, S. Ryu, A. Furusaki, and A. W. W. Ludwig, \textit{Phys. Rev. B} \textbf{78}, 195125 (2008).

\bibitem{Zhang1} X. - L. Qi, T. L. Hughes, S. Raghu, S. - C. Zhang, \textit{Phys. Rev. Lett.} \textbf{102}, 187001 (2009).

\bibitem{Zhang2} X. - L. Qi and S. - C. Zhang, \textit{Rev. Mod. Phys.} \textbf{83}, 1057 (2011).





\bibitem{Tinkham} M. Tinkham, Introduction to Superconductivity \textit{McGraw-Hill, New York,} (1996).

\bibitem{Liu} Z. Liu, X. Yao, J. Shao, M. Zuo, L. Pi, S. Tan, C. Zhang, and Y. Zhang, \textit{J. Am. Chem. Soc.} \textbf{137}, 10512–10515, (2015).

\bibitem{Bi2Se3_Zhang} H. Zhang, C.-X. Liu, X.-L. Qi, X. Dai, Z. Fang and S.-C. Zhang, \textit{Nat. Phys.}, \textbf{5}, 438, (2009).

\bibitem{Bi2Se3_Hasan} Y. Xia, D. Qian, D. Hsieh, L. Wray, A. Pal, H. Lin, A. Bansil, D. Grauer, Y. S. Hor, R. J. Cava and M. Z. Hasan, \textit{Nat. Phys.}, \textbf{5}, 398, (2009).

\bibitem{CuBi2Se3_Cava} Y. S. Hor, A. J. Williams, J. G. Checkelsky, P. Roushan, J. Seo, Q. Xu, H. W. Zandbergen, A. Yazdani, N. P. Ong, and R. J. Cava, \textit{Phys. Rev. Lett.} \textbf{104}, 057001, (2010).

\bibitem{CuBi2Se3_Ando} M. Kriener, K. Segawa, Z. Ren, S. Sasaki, and Y. Ando, \textit{Phys. Rev. Lett.} \textbf{106}, 127004, (2011).

\bibitem{SrBi2Se3_Pneha} Shruti, V. K. Maurya, P. Neha, P. Srivastava, and S. Patnaik, \textit{Phys. Rev. B} \textbf{92}, 020506(R), (2015).

\bibitem {MPSmiley2} M. P. Smylie, H. Claus, U. Welp, W.-K. Kwok, Y. Qiu, Y. S. Hor, and A. Snezhko, \textit{Phys. Rev. B} \textbf{94}, 180510(R), (2016).

\bibitem{SrBi2Se3_Han} C. Q. Han, \ et. al., \textit{App. Phys. Lett.} \textbf{107}, 171602, (2015).

\bibitem{CuBi2Se3_Levy_STM} N. Levy, T. Zhang, J. Ha, F. Sharifi, A. A. Talin, Y. Kuk, and J. A. Stroscio, \textit{Phys. Rev. Lett.} \textbf{110}, 117001,(2013).

\bibitem{SrBi2Se3_STM} G. Du, \ et. al., \textit{Nat. Comm.} \textbf{8}, 14466 (2017).

\bibitem{BCS} J. Bardeen, L. N. Cooper, and J. R. Schrieffer, \textit{Phys. Rev.} \textbf{108}, 1175 (1957).



\bibitem{CuBi2Se3_TSC1} S. Sasaki, M. Kriener, K. Segawa, K. Yada, Y. Tanaka, M. Sato, and Y. Ando, \textit{Phys. Rev. Lett.} \textbf{107}, 217001, (2011).

\bibitem{CuBi2Se3_TSC2} X. Chen, C. Huan, Y. S. Hor, C. A. R. Sá de Melo and Z. Jiang, \textit{arXiv:1210.6054v1} (2012).

\bibitem{CuBi2Se3_TSC3} T. H. Hsieh and L. Fu, \textit{Phys. Rev. Lett.} \textbf{108}, 107005, (2012).

\bibitem{CuBi2Se3_TSC4} T. Kirzhner, E. Lahoud, K. B. Chaska, Z. Salman, and A. Kanigel, \textit{Phys. Rev. B} \textbf{86}, 064517, (2012).




\bibitem{NBS_Qiu} Y. Qiu, K. N. Sanders, J. Dai, J. E. Medvedeva, W. Wu, P. Ghaemi, T. Vojta, Y. S. Hor, \textit{arXiv:1512.03519v1} (2015).

\bibitem{NBS_Asaba} T. Asaba, B. J. Lawson, C. Tinsman, L. Chen, P. Corbae, G. Li, Y. Qiu, Y. S. Hor, L. Fu, and L. Li
\textit{Phys. Rev. X} \textbf{7}, 011009 (2017).

\bibitem{NBS_Shen} J. Shen, W.-Y. He, N. F. Q. Yuan, Z. Huang, C.-w. Cho, S. H. Lee, Y. S. Hor, K. T. Law and R. Lortz, \textit{arXiv:1711.03265v2}  (2017).

\bibitem{NBS_Lawson} B. J. Lawson, P. Corbae, G. Li, F. Yu, T. Asaba, C. Tinsman, Y. Qiu, J. E. Medvedeva, Y. S. Hor, and L. Li, \textit{Phys. Rev. B} \textbf{94}, 041114(R), (2016).

\bibitem{MPSmiley1} M. P. Smylie, \ et. al., \textit{Phys. Rev. B} \textbf{96}, 115145, (2017).







\bibitem{BSTS} T. Knispel, W. Jolie, N. Borgwardt, J. Lux, Zhiwei Wang, Yoichi Ando, A. Rosch, T. Michely, and M. Grüninger, \textit{Phys. Rev. B} \textbf{96}, 195135, (2017).

\bibitem{Dynes} R. C. Dynes, V. Narayanamurti, and J. p. Garno, \textit{Phys. Rev. lett.} \textbf{41}, 1509 (1978).

\bibitem{Chiralpwave_Tanaka1} T. Yokoyama, C. Iniotakis, Y. Tanaka, and M. Sigrist, \textit{Phys. Rev. Lett.}, \textbf{100}, 177002, (2008).

\bibitem{Chiralpwave_Tanaka2} Y. Asano, A. A. Golubov, Y. V. Fominov, and Y. Tanaka, \textit{Phys. Rev. Lett}, \textbf{107}, 087001, (2011).




\end{thebibliography}
\end{document}